\documentclass{aa}
\usepackage{graphics,epsfig,lscape,times}

\usepackage{natbib}
\bibpunct{(}{)}{;}{a}{}{,}

\newcommand{\ergcm}[1]{$\times 10^{#1}$ erg cm$^{-2}$ s$^{-1}$}

\newcommand{\hcm}[1]{$\times 10^{#1}$ cm$^{-2}$}

\newcommand{\expo}[1]{$\times 10^{#1}$}

\newcommand{\nh}{$\mathrm{N_H}$}

\newcommand{\eqw}{\hbox{EW}}

\newcommand{\ltsima}{$\buildrel < \over \sim$}
\newcommand{\lsim}{\lower.5ex\hbox{\ltsima}}
\newcommand{\gtsima}{$\buildrel > \over \sim$}
\newcommand{\gsim}{\lower.5ex\hbox{\gtsima}}

\newcommand{\rxb}{\hbox{\object{RX\,J0720.4$-$3125}}}

\newcommand{\rbs}{\hbox{\object{RBS1223}}}

\begin{document}
 
\title{Evidence for precession of the isolated neutron star \object{RX\,J0720.4$-$3125}
        \thanks{Based on observations with XMM-Newton,
               an ESA Science Mission with instruments and contributions 
               directly funded by ESA Member states and the USA (NASA)}}
 
\author{F.~Haberl\inst{1} \and R.~Turolla\inst{2} \and C.P.~De Vries\inst{3} \and
        S.~Zane\inst{4} \and J.~Vink\inst{5} \and M.~M\'endez\inst{3} \and 
	F.~Verbunt\inst{5}}

\titlerunning{Precession of the isolated neutron star RX\,J0720.4$-$3125}
\authorrunning{Haberl et al.}
 
\offprints{F. Haberl, \email{fwh@mpe.mpg.de}}
 
\institute{Max-Planck-Institut f\"ur extraterrestrische Physik,
     Giessenbachstra{\ss}e, D-85748 Garching, Germany;
     \and
     Department of Physics, University of Padua, via Marzolo 8,
     I-35131, Padova, Italy; 
     \and
     SRON, Netherlands Institute for Space Research, Sorbonnelaan 2,
     NL-3584 CA Utrecht, The Netherlands; 
     \and
      Mullard Space Science Laboratory, University College London, Holmbury St. Mary,
     Dorking, Surrey, RH5 6NT, UK; 
     \and
     University Utrecht, PO Box 80000,
     NL-3508 TA Utrecht, The Netherlands; 
     }

\date{Received 27 February 2006 / Accepted 23 March 2006}
 
\abstract{
The XMM-Newton spectra of the isolated neutron star RX\,J0720.4$-$3125 obtained over 
4.5 years can be described by sinusoidal variations in the inferred blackbody 
temperature, the size of the emitting area and the depth of the absorption line with a 
period of 7.1$\pm$0.5 years, which we suggest to be the precession period of 
the neutron star. Precession of a neutron star with two hot spots of 
different temperature and size, probably not located exactly in antipodal 
positions, may account for the variations in the X-ray spectra, changes in the 
pulsed fraction, shape of the light curve and the phase-lag between soft 
and hard energy bands observed from RX\,J0720.4$-$3125. An independent 
sinusoidal fit to published and new pulse timing residuals from a coherent analysis 
covering $\sim$12 years yields a consistent period of 7.7$\pm$0.6 years supporting 
the precession model.

}
 
\maketitle
 
\section{Introduction}

The 8.39 s X-ray pulsar \rxb\ was discovered in the ROSAT 
all-sky survey data by
\citet{1997A&A...326..662H} and identified with a faint blue optical star
\citep{1998A&A...333L..59M,1998ApJ...507L..49K,2003ApJ...590.1008K}
which shows a proper motion of about 100 mas/yr \citep{2003A&A...408..323M}. 
The bright, soft X-ray source belongs to a small group of nearby radio-quiet isolated
neutron stars with blackbody-like thermal X-ray spectra 
\citep[for recent reviews
see][]{2000PASP..112..297T,2001xase.conf..244M,2004AdSpR..33..638H,2005fysx.conf...39H}.

Broad absorption lines in the X-ray spectra of most of these objects 
were reported and are usually interpreted as due to resonant absorption at 
the proton cyclotron energy and/or bound-bound transitions in H or H-like He
\citep[][ hereafter H04]{2003A&A...403L..19H,2004ApJ...608..432V,2005ApJ...627..397Z,2004A&A...419.1077H}.
This is suggestive of neutron star magnetic field strengths of 
$\approx$10$^{13}$--$10^{14}$~G, values which are consistent with those derived
from pulse timing for the period derivatives of \rxb\ 
\citep[][ hereafter K05]{2002MNRAS.334..345Z,2004MNRAS.351.1099C,2005ApJ...628L..45K} 
and \rbs\ \citep{2005ApJ...635L..65K} if one assumes magnetic dipole braking. Moreover, 
\rxb\ shows variations of the absorption line depth with pulse phase (H04) with the 
line being weakest near intensity maximum and deepest at the declining part of the pulse.

Among the seven thermal isolated neutron stars discovered with ROSAT,
\rxb\ is unique by showing a gradual change of the X-ray spectrum on
a time scale of years accompanied by an energy-dependent change in
the pulse profile \citep[][ hereafter D04]{2004A&A...415L..31D}.
In this letter we report on further XMM-Newton observations of this enigmatic 
source which show that the long-term trends have reversed. We present 
pulse-phase averaged and pulse-phase resolved spectra from the EPIC-pn 
instrument. The results of our spectral and temporal analysis show strong 
evidence for a cyclic variation of the X-ray properties which strengthens the 
case of a precessing neutron star.

\section{XMM-Newton observations}

\begin{table}
\caption[]{XMM-Newton EPIC-pn observations of \rxb.}
\begin{tabular}{lllll}
\hline\noalign{\smallskip}
\multicolumn{1}{l}{Orbit} &
\multicolumn{1}{l}{Observation} &
\multicolumn{1}{l}{Date} &
\multicolumn{1}{l}{Setup$^1$} &
\multicolumn{1}{l}{Exp. [s]} \\

\noalign{\smallskip}\hline\noalign{\smallskip}
0078 & 0124100101 & 2000 May 13     & FF thin   & 44275 \\
0175 & 0132520301 & 2000 Nov. 21-22 & FF medium & 22952 \\
0533 & 0156960201 & 2002 Nov. 6-7   & FF thin   & 25696 \\
0534 & 0156960401 & 2002 Nov. 8-9   & FF thin   & 27273 \\
0622 & 0158360201 & 2003 May 2-3    & SW thick  & 51022 \\
0711 & 0161960201 & 2003 Oct. 27    & SW thin   & 12711 \\
0711 & 0161960201 & 2003 Oct. 27-28 & SW medium & 17429 \\
0815 & 0164560501 & 2004 May 22-23  & FF thin   & 26642 \\
0986 & 0300520201 & 2005 April 28   & FF thin   & 37101 \\
1060 & 0300520301 & 2005 Sep. 23    & FF thin   & 34989 \\
1086 & 0311590101 & 2005 Nov. 12-13 & FF thin   & 33777 \\
\noalign{\smallskip}\hline
\end{tabular}

$^{(1)}$ Read-out mode and filter; FF: Full Frame;  
         SW: Small Window. 
\label{xmm-obs}
\end{table}

The soft X-ray source \rxb\ was observed with XMM-Newton \citep{2001A&A...365L...1J} as 
calibration target until 2004 when it was recognised to exhibit a variable spectrum.
Since then we monitored the spectral evolution of the source. 
Here we utilise the data collected with the EPIC-pn camera \citep[][]{2001A&A...365L..18S}.
The details of the XMM-Newton observations with the instrumental setup used for
EPIC-pn are summarised in Table~\ref{xmm-obs}. \rxb\ is slightly too bright
for the full-frame (FF) read-out mode causing photon pile-up effects which systematically
harden the spectrum (for details of this effect see H04).
For a relative comparison of the spectra we ignore this effect but emphasise
that small systematic differences are present between the FF mode spectra and those obtained
in the faster small window (SW) mode which are free of pile-up effects. 
We do not use the observation with thick filter due to the much lower
efficiency and different detector response.
We screened out strong background flares and list the resulting net exposure times 
(including dead time which is important for SW mode) in Table~\ref{xmm-obs}.
We processed the data using the XMM-Newton analysis system SAS6.5 and extracted spectra
and light curves from circular regions with radius of 30\arcsec.

\subsection{Pulse-phase averaged spectra}

\begin{figure}
\begin{center}
\resizebox{\hsize}{!}{\includegraphics[clip=,angle=-90]{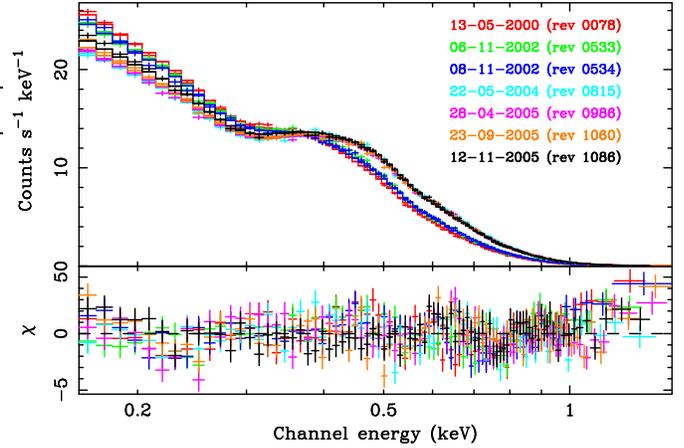}}
\end{center}
\caption{Pulse-phase averaged EPIC-pn spectra of \rxb\ from the seven observations with
   the same instrumental setup (FF read-out mode, thin filter). The same detector 
   efficiency for all the spectra allows a direct comparison and demonstrates the 
   long-term spectral changes. The softest spectrum (the uppermost at low
   energies) was obtained in May 2000, while the hardest (the lowest at
   low energies) is that from May 2004. After May 2004 the spectra 
   increased monotonically at low energies.}
\label{fig-spectra}
\end{figure}

We extracted and analyzed pulse-phase averaged spectra for the 10 observations 
(at different epochs and different instrumental setup) following H04 who found that
an absorbed blackbody model with a broad Gaussian absorption line represents the 
spectra of \rxb. We adopted this model and fitted the spectra simultaneously with
the absorption column density (\nh), the line energy and width common to all spectra.
We allowed temperature and the normalizations of blackbody and line to vary freely
between the spectra (except when observations were performed shortly after each other
as those from satellite revolutions 533/534 and the two parts of 711).
We present the EPIC-pn spectra obtained in FF mode with thin filter together with their 
best-fit model in Fig.~\ref{fig-spectra} which shows that the long-term trend of spectral
hardening has reversed. We report the fit results in Table~\ref{tab-fits}. The common 
parameters obtained from the best fit are \nh\ = 1.01$\pm$0.03 \hcm{20}, a line energy 
of 280$\pm$6~eV with a width of $\sigma = 90\pm$5~eV. 

\begin{figure}
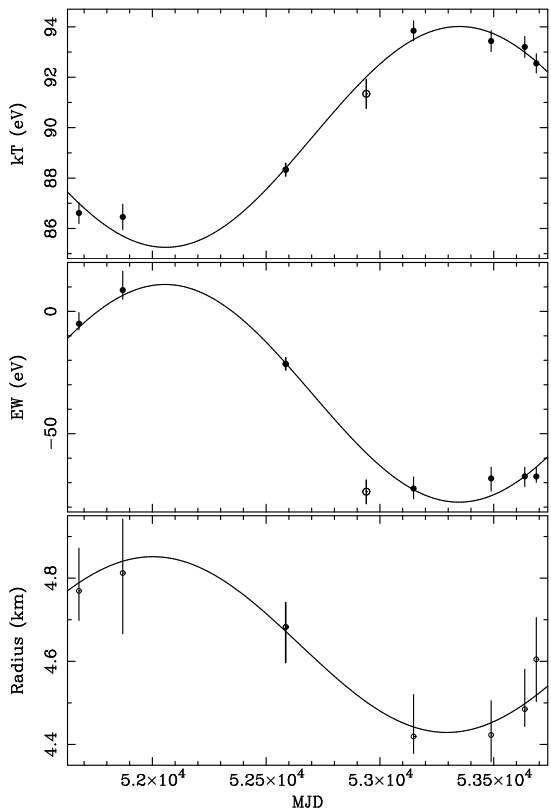

\begin{center}
\resizebox{7.19cm}{!}{\includegraphics[clip=,angle=-90]{allthinmed_kT.ps}}
\resizebox{7.19cm}{!}{\includegraphics[clip=,angle=-90]{allthinmed_eqw.ps}}
\resizebox{7.19cm}{!}{\includegraphics[clip=,angle=-90]{allthinmed_rns.ps}}
\end{center}
\caption{Variation of temperature, line equivalent width and radius of the emitting region
   of \rxb\ derived from a simultaneous fit to the EPIC-pn spectra using an absorbed 
   blackbody model with a broad absorption line at 280 eV. The sine wave with a period 
   of 7.1 years indicates the best fit to the data derived from the FF mode (marked with 
   filled circle) observations. For the sine fit of EW and radius we fixed the period at 
   the value derived from kT. SW mode data from Oct. 2003 (open circle) are not used in 
   the fits.}
\label{fig-precession}
\end{figure}

\begin{table}[t]
\caption[]{Spectral analysis of pulse-phase averaged X-ray spectra.}
\begin{tabular}{lccc}
\hline\noalign{\smallskip}
\multicolumn{1}{l}{Orbit} &
\multicolumn{1}{c}{kT [eV]} &
\multicolumn{1}{c}{\eqw\ [eV]} &
\multicolumn{1}{c}{Flux$^{(1)}$ [erg cm$^{-2}$]} \\

\noalign{\smallskip}\hline\noalign{\smallskip}
0078     & 86.6$\pm$0.4 &  $-$5.02$\pm$4.5  & 0.98\expo{-11} \\
0175     & 86.5$\pm$0.5 &  $+$8.68$\pm$7.7  & 1.01\expo{-11} \\
0533/534 & 88.3$\pm$0.3 &  $-$21.5$\pm$2.6  & 0.98\expo{-11} \\
0711/711 & 91.3$\pm$0.6 &  $-$73.7$\pm$4.9  & 1.19\expo{-11} \\
0815     & 93.8$\pm$0.4 &  $-$72.4$\pm$4.7  & 1.02\expo{-11} \\
0986     & 93.5$\pm$0.4 &  $-$68.3$\pm$5.2  & 1.02\expo{-11} \\
1060     & 93.2$\pm$0.4 &  $-$67.4$\pm$4.3  & 1.02\expo{-11} \\
1086     & 92.6$\pm$0.4 &  $-$67.5$\pm$3.5  & 1.04\expo{-11} \\
\noalign{\smallskip}\hline\noalign{\smallskip}
\end{tabular}

Throughout the paper errors are given for a 90\% confidence level. \\
$^{(1)}$Observed flux is given in the energy band 0.1$-$2.4 keV. The
systematically higher flux from satellite revolution 711 is caused
by the systematic difference of FF and SW mode (FF pile-up losses).
The statistical uncertainties on the flux are less then 2\ergcm{-14}, which are 
negligible compared to the stability of the instrument of about 1\% between
observations. 
\label{tab-fits}
\end{table}

The best fit values for the blackbody temperature (kT), the line equivalent width
(EW) and the inferred emitting area (radius for a circular region with an assumed source 
distance of 300 pc) as function of time are shown in Fig.~\ref{fig-precession}. 
All parameters can be described with a sinusoidal variation and we derive a period 
of 2580$\pm$180 days for the variation in temperature. In the fit we do not 
include the values obtained from the SW mode spectra because of the possible systematic 
difference (pile-up, see above).
Among the three parameters the temperature can be probably best constrained from the 
X-ray spectra. It is also not clear if there are additional variations in the
absorption line energy and width which we assume to be constant in our current analysis.
Therefore, we also fit a sine to the variations in EW and emission radius, but fix the
period at the value found from the temperature variation. 

\subsection{Pulse-phase variations}

\begin{figure}
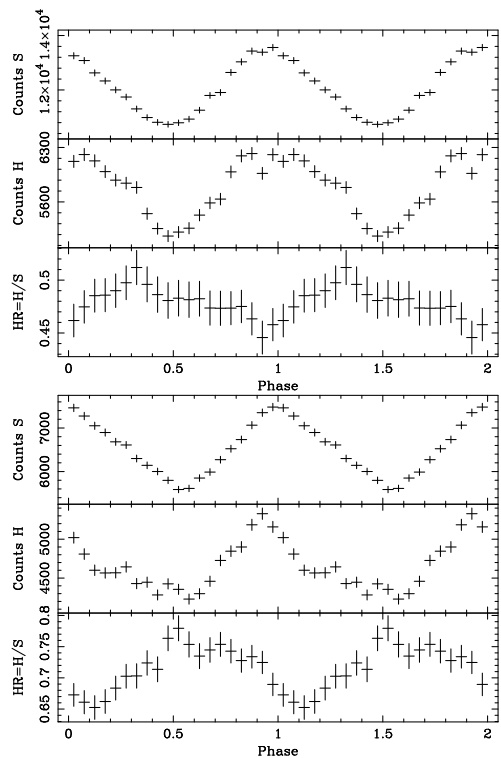

\begin{center}
\resizebox{6.5cm}{!}{\includegraphics[clip=,angle=-90]{P0124100101PNS003PIEVLI0000_sc_hr_phase.ps}}
\vspace{3mm}
\resizebox{6.5cm}{!}{\includegraphics[clip=,angle=-90]{P0164560501PNS001PIEVLI0000_sc_hr_phase.ps}}
\end{center}
\caption{Pulse profiles of \rxb\ in two different energy bands (soft S: 0.12-0.4 keV; 
   hard H: 0.4-1.0 keV) together with the hardness ratio HR=H/S. The top panels are 
   obtained from the observation in May 2000 and the bottom panels from that of May 2004.}
\label{fig-profile}
\end{figure}

We assigned pulse phases to all detected events using the X-ray timing 
ephemerides inferred by K05 (the ``All Data'' solution)
and folded the light curves in different energy bands to produce pulse profiles.
Examples from two XMM-Newton observations are presented in Fig.~\ref{fig-profile}.
As was found by D04 the pulse profile became deeper with time. In 2004 
the shape of the profile changed in particular in the hard band and the hardness ratio was 
higher on average and showed more modulation as compared to the observation in May 2000. 
To investigate the spectral evolution as function of pulse phase 
we divided the pulse into five phase intervals of equal length with the first interval from 
phase 0.0 to 0.2.

\begin{figure}
\begin{center}
\resizebox{8.1cm}{!}{\includegraphics[clip=,angle=-90]{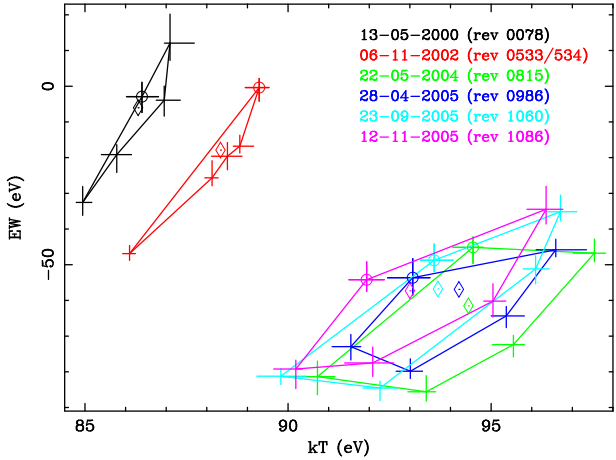}}
\end{center}
\caption{Equivalent width of the absorption line vs. temperature kT derived from
   the FF observations with thin filter. Lozenges denote the values derived 
   from the phase-averaged spectra. During the pulse the parameters evolve 
   counter-clockwise, the circle marks phase 0.0-0.2.}
\label{fig-pulse}
\end{figure}

To avoid any systematic shifts due to different instrumental setups, we use 
only the seven observations in FF mode with thin filter. Similarly to the 
analysis of phase-averaged spectra, we performed a simultaneous fit with the 
same model to the 5$\times$7 spectra. \nh, the line 
energy and width were treated common to all spectra and were found to be 
consistent within the errors with the values derived from the phase-averaged
spectra. In Fig.~\ref{fig-pulse} the derived line EW is plotted versus 
temperature kT. For each observation the evolution of the two parameters 
during the X-ray pulse follows an ellipse-like track (sampled by five points 
from our finite number of phase intervals) in the kT-EW plane. The evolution 
proceeds counter-clockwise; the point marked with a circle indicates the 
phase interval 0.0-0.2. Several remarkable features are seen in 
Fig.~\ref{fig-pulse}: 1) The variation in kT was smaller during the 
first observations, consistent with the $\sim$2.5~eV value reported by H04.
During the later observations the amplitude in the kT variation increased 
to $\sim$6~eV, almost as large as the long-term change of $\sim$8~eV seen 
in the phase-averaged spectra (Fig.~\ref{fig-precession}). 2) The 
amplitude in the line EW variation is $\sim$40~eV and did not change 
significantly between the observations.
3) The long-term 
trend reversal of the evolution seen from the phase-averaged spectra 
(Fig.~\ref{fig-precession}) after the May 2004 observation is similarly 
seen at all pulse phases. We note that during the 
pulse variation of the first observation the line is formally detected as
emission line in phase interval 0.8-1.0, i.e. before reaching the 
intensity maximum. Also during the second observation there is an 
indication that the line in the phase-averaged spectrum is in emission 
rather than absorption. However, we can not exclude that this is caused
by uncertainties in the calibration which may result in a systematic shift of the 
EW by $\sim$10~eV (for all spectra). 

\section{Discussion}

The medium resolution X-ray spectra obtained by the  
EPIC-pn instrument show that the long-term spectral evolution 
discovered by D04 has reversed. The spectra 
became softer following the observation of May 2004 when the 
phase-averaged spectrum was hardest. All spectra can be well modeled
by an absorbed blackbody continuum with a broad absorption line of 
Gaussian shape. The spectral variations both on time scales of years
and during the 8.39 s pulse can be explained by variations in the inferred
blackbody temperature, the depth of the absorption line and the size of the
emitting area.
In the simplest approach we did not allow changes in the energy and 
width of the line. If the line is caused by cyclotron absorption 
both parameters may be expected to vary. This requires a more 
detailed investigation of the spectra also involving the higher resolution
RGS data which is work in progress.

Blackbody temperature, size of emitting area and absorption line depth 
(equivalent width) derived from the phase-averaged 
spectra follow a sinusoidal variation with a period of ~7.1 years. 
Although the XMM-Newton observations do not yet cover a full period,
it should be noted that the temperature of 79$\pm$4~eV derived from the 
ROSAT  spectrum (H04), obtained in Sept. 1993, $\sim$6.6 years before the first 
XMM-Newton observation is also consistent with a periodic behavior of \rxb.
The involved time scale is strongly suggestive of a scenario with a 
freely precessing neutron star, as first suggested by D04. The 
observed period of $\sim$7.1 yr is then naturally interpreted as the precession 
period of the neutron star. 

In that case the precession period should also be visible in the residual 
phase shifts seen in the coherent timing analysis of the spin evolution of 
\rxb\ (K05). Therefore, we fitted the phase residuals as inferred by K05
with a sinusoidal function, including the three EPIC-pn data points for 
the new XMM-Newton observations of 2005. 
The additional data points formally exclude the cubic model used by K05 
and favour the sinusoidal model. We obtain a 
preliminary period of 2830$\pm$220 days (7.7$\pm$0.6 years) from the phase 
residual analysis. This period is somewhat longer, but consistent with that
inferred from the spectral analysis. The three ROSAT observations extend 
the time coverage to more than 12 years, i.e. more than 1.5 precession 
cycles. The results of a detailed timing analysis which directly includes 
precession in the spin frequency model will be presented in a forthcoming 
paper. 

A picture based on thermal emission from two hot spots on the
surface of a freely precessing neutron stars appears indeed
promising in explaining many of the peculiar characteristics of 
\rxb\ which have been a challenge so far. In this scenario
the long-term change in temperature is produced by the different
(phase-averaged) fractions of the two spots which enter into view
as the star precesses. In order for such a model to work, the two
emitting regions need to have different temperatures and sizes, as
it has been recently proposed in the case of another member of the
same class \citep[\rbs, ][]{2005A&A...441..597S}. 
The nearly, but not exactly sinusoidal pulse profile 
(Fig.~\ref{fig-profile}, D04) and the - to first order - successful 
fit of a phase-connected timing solution indicates that the poles can 
not be exactly antipodal, but must be predominantly in the same 
(East-West) hemisphere.

This is again similar 
to what was found for \rbs. During the first XMM-Newton observation (May
2000), $\mathrm{kT}\sim 85$~eV and the cooler, larger spot was
predominantly visible, while four years later (May 2004),
precession brought into view mostly the second, hotter and smaller spot,
increasing the temperature to $\sim$95~eV. This may explain the
observed temperature/emitting area variations and their
anti-correlation (see Fig.~\ref{fig-precession}). It should also be noted
here that the total flux we see from the source stays quite constant, i.e.
the two poles contribute about the same amount. The presence of two different,
non-antipodal spots may also account for the observed changes in
the pulsed fraction, light curve shape and for the apparent (spin)
phase-lag seen between the soft and hard band.

\begin{figure*}
\begin{center}
\resizebox{7.8cm}{!}{\includegraphics[clip=]{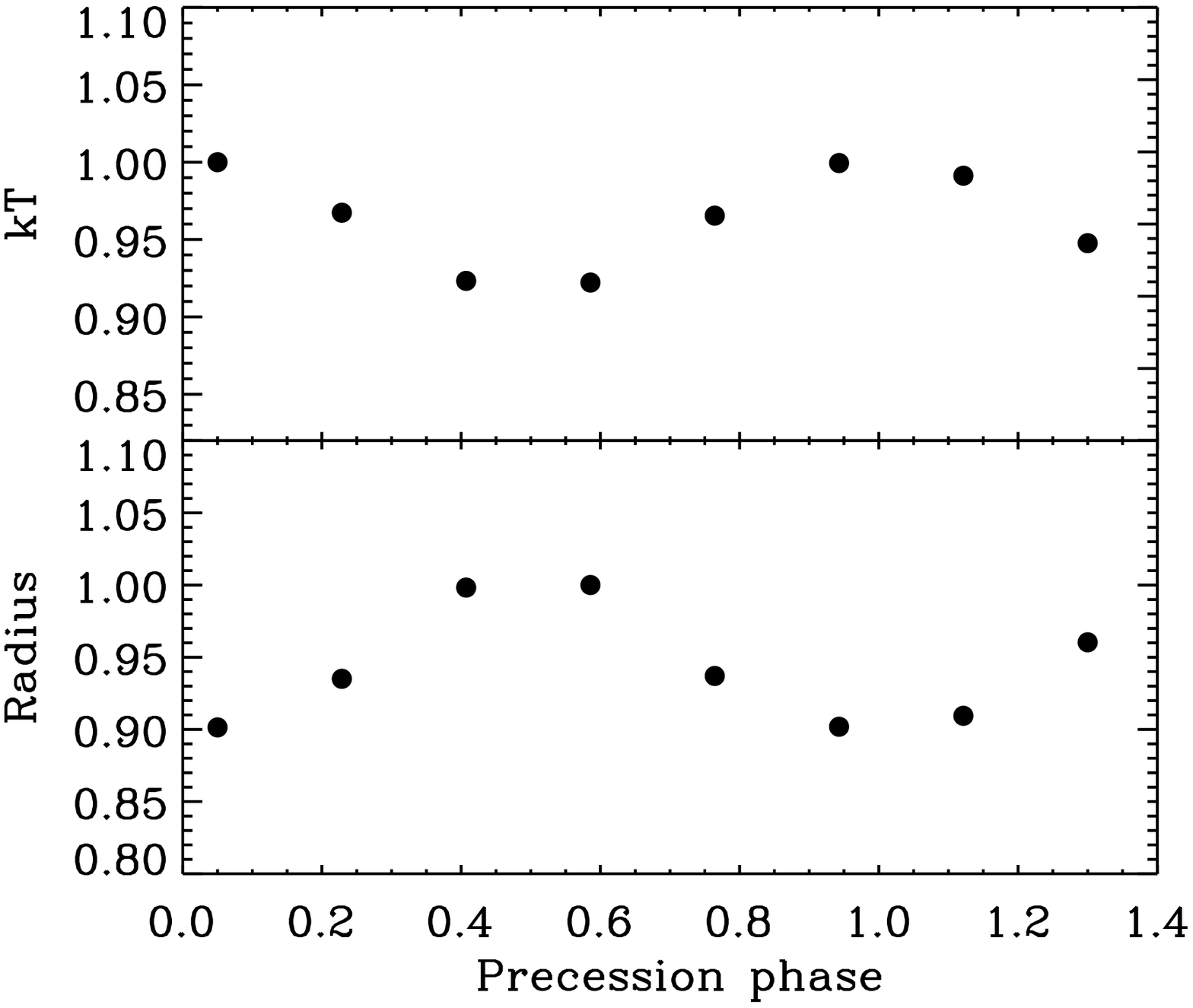}}
\resizebox{7.8cm}{!}{\includegraphics[clip=]{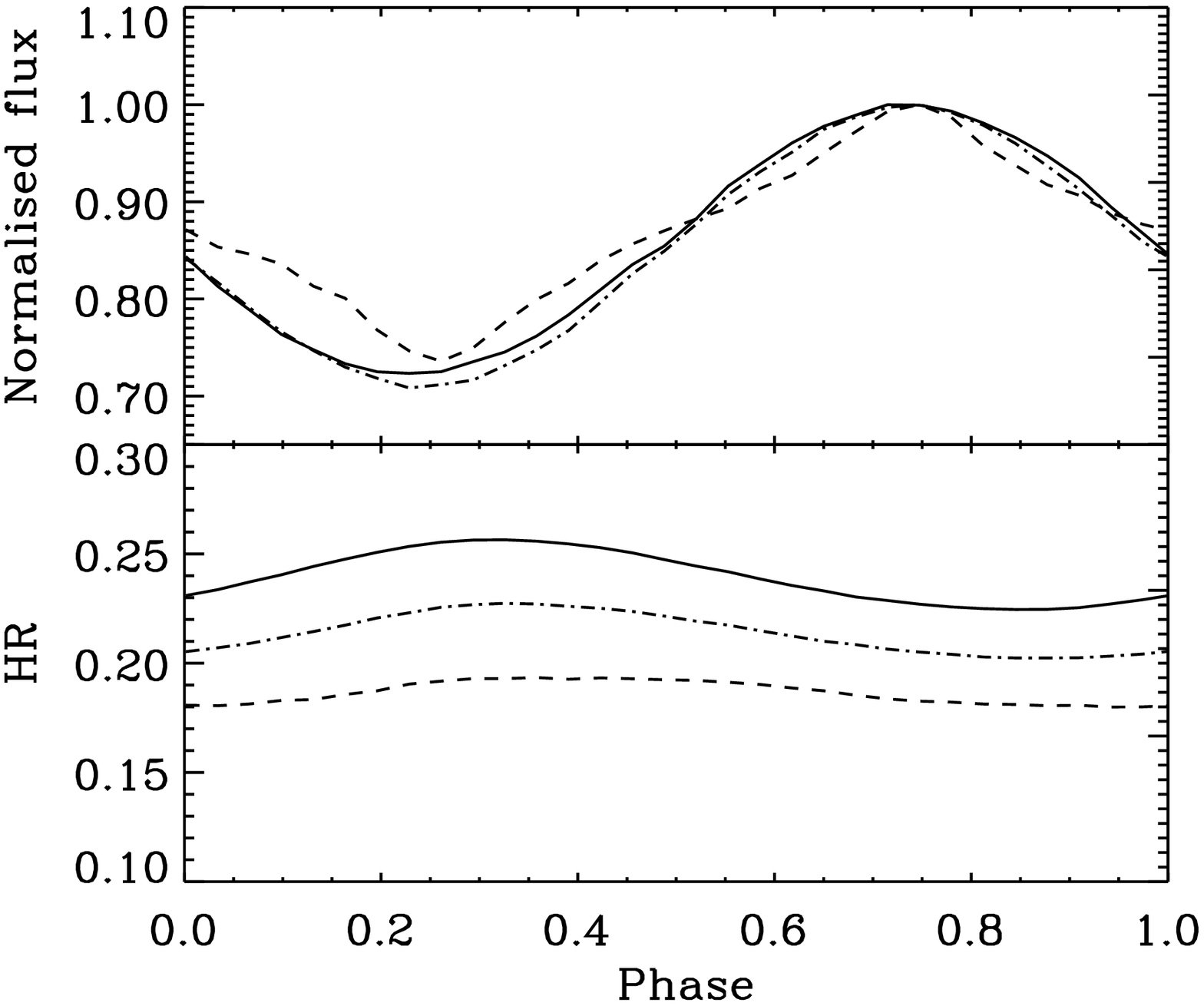}}
\end{center}
\caption[]{Left panel: the variation of the blackbody temperature and radius
   (both normalised) as a function of the precession phase. Right panel: the 
   evolution of the light curve (0.12$-$1~keV band) and hardness ratio with 
   precession phase; full, dashed and dash-dotted lines refer to phases 
   0.05, 0.4 and 0.75, respectively. Here $\alpha=20^\circ$, 
   $\mathrm{T_1}=80$~eV, $\mathrm{T_2}=100$~eV, $\sin\theta_1=0.8$, 
   $\sin\theta_2=0.6$, $\theta_0=160^\circ$, $\chi=75^\circ$ and $\xi=5^\circ$.}
 \label{figmod}
\end{figure*}

In Fig.~\ref{figmod} we show the results obtained for such a model with
a numerical code adapted from that discussed by \citet{2006MNRAS.366..727Z}. 
First the phase-resolved spectrum is computed, having chosen values 
of the temperature and size of the two spots ($\mathrm{T_{1,2}}$, 
$\theta_{1,2}$), and of their relative angular displacement 
$\theta_0$, to match the main observational features of 
\rxb. We then compute the phase-averaged 
spectrum and repeat the calculation at different precession phases for a chosen
value of the precession angle $\alpha$. The two angles $\chi$ and $\xi$ denote
the inclination of the line-of-sight 
with respect to the precession axis and that between the centre of one spot and 
the rotation axis, respectively. We assume isotropic blackbody emission from 
the caps and no attempt has been made to model the absorption feature. 
Relativistic ray-bending is included. Since the present calculation has mainly 
illustrative purposes, we do not account for the detector response and 
interstellar absorption nor do we look for parameter fine-tuning. 

As can be seen from Fig.~\ref{figmod}, the model is capable of reproducing the 
observed variations of the blackbody temperature and emitting area, and 
also their phase anti-correlation. The anti-correlation of the hardness ratio 
with the total intensity is also recovered, and the maximum of the hardness 
ratio shifts ahead in phase. The light curve modulation evolves and the shape 
changes, becoming more sharply peaked at certain epochs.
Although no proper fitting of the observed light curves and spectra has been 
attempted at this stage (work in 
progress), our preliminary analysis strongly supports the idea that \rxb\ 
is a precessing, nearly aligned rotator, seen almost equator-on. 
First indications seem to favour precession angles $\ga 10^\circ$, 
larger than those found for radio-pulsars.
We point out that the modulation is sensitive to the value of M/R 
(2GM/Rc$^2$ = 0.42 in our example of Fig.~\ref{figmod}). Although present 
uncertainties prevent us from reaching any firm conclusion, future improvements in 
the modelling and new observations may allow us to derive  
constraints on the equation of state.

The variations we see in the absorption line could also be related to the different 
properties of the two emitting spots. Two misaligned caps are suggestive of a magnetic
field structure more complex than a core-centred dipole. A different field 
configuration could be then the origin of the different temperature and size 
of the spots. In both \rbs\ and \rxb\ the inferred size of the 
hotter spot is smaller than that of the cooler spot, probably because the former is 
more confined due to a steeper surface gradient of the magnetic field. A further 
possibility is that the line
changes (both with spin and precession phase) arise because of geometrical effects.
The rays which reach the observer are at different angles with the magnetic field at 
different phases. This means that the contribution of ordinary and extraordinary 
photons to the total spectrum is phase-dependent. Since extraordinary photons 
contribute most to resonant absorption at the proton cyclotron energy, one expects 
the line properties to change with phase. 

A detailed analysis of the implications of the precession of \rxb\ 
for the physics of neutron star interiors is outside 
the scope of this letter. We note, however, that the star distortion 
$\epsilon=(\mathrm{I}_3-\mathrm{I}_1)/\mathrm{I}_1=\mathrm{P_{spin}}/\mathrm{P_{prec}}\approx 4\times 10^{-8}$ 
is larger than that reported for radio-pulsars
\citep[e.g.][]{2001MNRAS.324..811J,2006MNRAS.365..653A} 
but smaller than that for Her\,X-1 \citep{2000Ketsaris}.

\begin{acknowledgements}
The XMM-Newton project is supported by the Bundesministerium f\"ur Bildung und
For\-schung / Deutsches Zentrum f\"ur Luft- und Raumfahrt (BMBF/DLR), the
Max-Planck-Gesellschaft and the Heidenhain-Stif\-tung. SZ acknowledges support 
from a PPARC AF. We thank Joachim Tr\"umper, Mark Cropper and Gianluca Israel 
for fruitful discussions.
\end{acknowledgements}

\bibliographystyle{aa}
\bibliography{ins,general,myrefereed,myunrefereed}

\begin{thebibliography}{26}
\expandafter\ifx\csname natexlab\endcsname\relax\def\natexlab#1{#1}\fi

\bibitem[{{Akg{\"u}n} {et~al.}(2006){Akg{\"u}n}, {Link}, \&
  {Wasserman}}]{2006MNRAS.365..653A}
{Akg{\"u}n}, T., {Link}, B., \& {Wasserman}, I. 2006, \mnras, 365, 653

\bibitem[{{Cropper} {et~al.}(2004){Cropper}, {Haberl}, {Zane}, \&
  {Zavlin}}]{2004MNRAS.351.1099C}
{Cropper}, M., {Haberl}, F., {Zane}, S., \& {Zavlin}, V.~E. 2004, \mnras, 351,
  1099

\bibitem[{{de Vries} {et~al.}(2004){de Vries}, {Vink}, {M{\'e}ndez}, \&
  {Verbunt}}]{2004A&A...415L..31D}
{de Vries}, C.~P., {Vink}, J., {M{\'e}ndez}, M., \& {Verbunt}, F. 2004, \aap,
  415, L31 (D04)

\bibitem[{{Haberl}(2004)}]{2004AdSpR..33..638H}
{Haberl}, F. 2004, Advances in Space Research, 33, 638

\bibitem[{{Haberl}(2005)}]{2005fysx.conf...39H}
{Haberl}, F. 2005, in 5 years of Science with XMM-Newton, 39

\bibitem[{{Haberl} {et~al.}(1997){Haberl}, {Motch}, {Buckley}, {Zickgraf}, \&
  {Pietsch}}]{1997A&A...326..662H}
{Haberl}, F., {Motch}, C., {Buckley}, D. A.~H., {Zickgraf}, F.~J., \&
  {Pietsch}, W. 1997, \aap, 326, 662

\bibitem[{{Haberl} {et~al.}(2003){Haberl}, {Schwope}, {Hambaryan}, {Hasinger},
  \& {Motch}}]{2003A&A...403L..19H}
{Haberl}, F., {Schwope}, A.~D., {Hambaryan}, V., {Hasinger}, G., \& {Motch}, C.
  2003, \aap, 403, L19

\bibitem[{{Haberl} {et~al.}(2004){Haberl}, {Zavlin}, {Tr{\"u}mper}, \&
  {Burwitz}}]{2004A&A...419.1077H}
{Haberl}, F., {Zavlin}, V.~E., {Tr{\"u}mper}, J., \& {Burwitz}, V. 2004, \aap,
  419, 1077 (H04)

\bibitem[{{Jansen} {et~al.}(2001){Jansen}, {Lumb}, {Altieri}, {Clavel}, {Ehle},
  {Erd}, {Gabriel}, {Guainazzi}, {Gondoin}, {Much}, {Munoz}, {Santos},
  {Schartel}, {Texier}, \& {Vacanti}}]{2001A&A...365L...1J}
{Jansen}, F., {Lumb}, D., {Altieri}, B., {et~al.} 2001, \aap, 365, L1

\bibitem[{{Jones} \& {Andersson}(2001)}]{2001MNRAS.324..811J}
{Jones}, D.~I. \& {Andersson}, N. 2001, \mnras, 324, 811

\bibitem[{{Kaplan} \& {van Kerkwijk}(2005{\natexlab{a}})}]{2005ApJ...628L..45K}
{Kaplan}, D.~L. \& {van Kerkwijk}, M.~H. 2005{\natexlab{a}}, \apjl, 628, L45
(K05)

\bibitem[{{Kaplan} \& {van Kerkwijk}(2005{\natexlab{b}})}]{2005ApJ...635L..65K}
{Kaplan}, D.~L. \& {van Kerkwijk}, M.~H. 2005{\natexlab{b}}, \apjl, 635, L65

\bibitem[{{Kaplan} {et~al.}(2003){Kaplan}, {van Kerkwijk}, {Marshall},
  {Jacoby}, {Kulkarni}, \& {Frail}}]{2003ApJ...590.1008K}
{Kaplan}, D.~L., {van Kerkwijk}, M.~H., {Marshall}, H.~L., {et~al.} 2003, \apj,
  590, 1008

\bibitem[{{Ketsaris} {et~al.}(2000){Ketsaris}, {Kuster}, {Postnov},
  {Prokhorov}, {Shakura}, {Staubert}, \& {Wilms}}]{2000Ketsaris}
{Ketsaris}, N.~A., {Kuster}, M., {Postnov}, K.~A., {et~al.} 2000, in {Proc.
  Intl. Workshop: Hot Points in Astrophysics, JINR, Dubna, Russia}, 192

\bibitem[{{Kulkarni} \& {van Kerkwijk}(1998)}]{1998ApJ...507L..49K}
{Kulkarni}, S.~R. \& {van Kerkwijk}, M.~H. 1998, \apjl, 507, L49

\bibitem[{{Motch}(2001)}]{2001xase.conf..244M}
{Motch}, C. 2001, in X-ray Astronomy, Stellar Endpoints, AGN, and the Diffuse
  X-ray Background, AIP Conference Proceedings, 244

\bibitem[{{Motch} \& {Haberl}(1998)}]{1998A&A...333L..59M}
{Motch}, C. \& {Haberl}, F. 1998, \aap, 333, L59

\bibitem[{{Motch} {et~al.}(2003){Motch}, {Zavlin}, \&
  {Haberl}}]{2003A&A...408..323M}
{Motch}, C., {Zavlin}, V.~E., \& {Haberl}, F. 2003, \aap, 408, 323

\bibitem[{{Schwope} {et~al.}(2005){Schwope}, {Hambaryan}, {Haberl}, \&
  {Motch}}]{2005A&A...441..597S}
{Schwope}, A.~D., {Hambaryan}, V., {Haberl}, F., \& {Motch}, C. 2005, \aap,
  441, 597

\bibitem[{{Str{\"u}der} {et~al.}(2001){Str{\"u}der}, {Briel}, {Dennerl},
  {Hartmann}, {Kendziorra}, {Meidinger}, {Pfeffermann}, {Reppin}, {Aschenbach},
  {Bornemann}, {Br{\"a}uninger}, {Burkert}, {Elender}, {Freyberg}, {Haberl},
  {Hartner}, {Heuschmann}, {Hippmann}, {Kastelic}, {Kemmer}, {Kettenring},
  {Kink}, {Krause}, {M{\"u}ller}, {Oppitz}, {Pietsch}, {Popp}, {Predehl},
  {Read}, {Stephan}, {St{\"o}tter}, {Tr{\"u}mper}, {Holl}, {Kemmer}, {Soltau},
  {St{\"o}tter}, {Weber}, {Weichert}, {von Zanthier}, {Carathanassis}, {Lutz},
  {Richter}, {Solc}, {B{\"o}ttcher}, {Kuster}, {Staubert}, {Abbey}, {Holland},
  {Turner}, {Balasini}, {Bignami}, {La Palombara}, {Villa}, {Buttler},
  {Gianini}, {Lain{\'e}}, {Lumb}, \& {Dhez}}]{2001A&A...365L..18S}
{Str{\"u}der}, L., {Briel}, U., {Dennerl}, K., {et~al.} 2001, \aap, 365, L18

\bibitem[{{Treves} {et~al.}(2000){Treves}, {Turolla}, {Zane}, \&
  {Colpi}}]{2000PASP..112..297T}
{Treves}, A., {Turolla}, R., {Zane}, S., \& {Colpi}, M. 2000, \pasp, 112, 297


\bibitem[{{van Kerkwijk} {et~al.}(2004){van Kerkwijk}, {Kaplan}, {Durant},
  {Kulkarni}, \& {Paerels}}]{2004ApJ...608..432V}
{van Kerkwijk}, M.~H., {Kaplan}, D.~L., {Durant}, M., {Kulkarni}, S.~R., \&
  {Paerels}, F. 2004, \apj, 608, 432

\bibitem[{{Zane} {et~al.}(2005){Zane}, {Cropper}, {Turolla}, {Zampieri},
  {Chieregato}, {Drake}, \& {Treves}}]{2005ApJ...627..397Z}
{Zane}, S., {Cropper}, M., {Turolla}, R., {et~al.} 2005, \apj, 627, 397

\bibitem[{{Zane} {et~al.}(2002){Zane}, {Haberl}, {Cropper}, {Zavlin}, {Lumb},
  {Sembay}, \& {Motch}}]{2002MNRAS.334..345Z}
{Zane}, S., {Haberl}, F., {Cropper}, M., {et~al.} 2002, \mnras, 334, 345

\bibitem[{{Zane} \& {Turolla}(2006)}]{2006MNRAS.366..727Z}
{Zane}, S. \& {Turolla}, R. 2006, \mnras, 366, 727

\end{thebibliography}

\end{document}